\documentclass[twocolumn,showpacs,pra,aps]{revtex4-1}
\usepackage{graphicx}
\usepackage{epstopdf}
\usepackage{amsmath}
\usepackage{amssymb}
\usepackage{epsfig}
\usepackage{amsthm}
\usepackage{color}
\usepackage{textcomp}

\begin{document}
\title{Long-distance temporal quantum ghost imaging over optical fibers}

\author{Shuai Dong}
\author{Wei Zhang}
\author{Yidong Huang}
\author{Jiangde Peng}
\affiliation{Tsinghua National Laboratory for Information Science and Technology, Department of Electronic Engineering, Tsinghua University, Beijing, 100084, China}

\affiliation{Corresponding author: zwei@tsinghua.edu.cn}

\begin{abstract}
Since the first quantum ghost imaging (QGI) experiment in 1995, many QGI schemes have been put forward. However, the position-position or momentum-momentum correlation required in these QGI schemes cannot be distributed over optical fibers, which limits their large geographical applications. In this paper, we propose and demonstrate a scheme for long distance QGI utilizing frequency correlated photon pairs. In this scheme, the frequency correlation is transformed to the correlation between the illuminating position of one photon and the arrival time of the other photon, by which QGI can be realized in the time domain. Since frequency correlation can be preserved when the photon pairs are distributed over optical fibers, this scheme provides a way to realize long-distance QGI over large geographical scale. In the experiment, long distance QGI over 50 km optical fibers has been demonstrated.
\end{abstract}

\maketitle

Quantum ghost imaging (QGI) [1] has attracted much attention in last two decades due to its abundance in physics and potential for quantum communication and quantum sensing. Originally, QGI was realized by photon pairs generated by spontaneous parametric down conversion (SPDC) in nonlinear crystals. The momentum-momentum correlation in the photon pairs was utilized in QGI to realize imaging in a nonlocal manner. After that, many ghost imaging schemes have been proposed [2-10]. The concept of the ghost imaging has been deeply developed ``from quantum to classical to computational'' [11]. While, the motivation of researches on ghost imaging also developed from the interest on fundamental physics [12,13] to applications such as information transmission and sensing [14]. It has been reported that ghost imaging can be used in optical information encryption and transmission [14]. However, it is still a problem how to realize QGI over large geographical scale by techniques compatible with optical communication and networks, since single mode fiber (SMF) used in optical communication does not support the distribution of the position-position or momentum-momentum correlation required in previous QGI schemes. 

In this paper, we  propose and demonstrate a scheme realizing long-distance QGI over optical fibers, based on frequency correlated photon pairs. It's well known that the frequency is a stable degree of freedom (DOF) of photons traveling in optical fibers. The frequency correlation can be preserved when the photon pairs are distributed over optical fibers. After long distance distribution over optical fibers, the two photons in the pair are sent to Alice and Bob sides, respectively. At Alice side, the photons are spatially dispersed to different directions according to their frequencies by a spatial dispersion component, such as a grating, then illuminate the object on different positions along a line. At Bob side, a component with large temporal dispersion is used to change the arrival time of idler photons with different frequencies when detected by a single photon detector (SPD). Hence, the frequency correlation in the initial photon pairs is transformed to the correlation of the illuminating positions of the photons at Alice side and the arrival time of the photons at Bob Side. Based on this correlation, the image along the illuminating line on the object can be obtained in time domain by the coincidence measurement, realizing one-dimensional QGI. Two-dimensional imaging can also be realized by step-moving the object, realizing a function of long distance ``quantum fax machine'' over optical fibers.

Frequency correlated photon pairs can be generated in third order nonlinear waveguide by the spontaneous four wave mixing process (SFWM)[15,16]. Their state can be expressed as
{\setlength\arraycolsep{2pt}
\begin{equation}
\vert\Psi\rangle=\int\mathrm{d}\Omega f(\Omega)\vert\omega_p+\Omega\rangle_s\vert\omega_p-\Omega\rangle_i,
\end{equation}
where ${\omega }_{p}$ is the frequency of the pump light. The indices $s$ and $i$ indicate the signal photons (with frequency ${\omega }_{s}$) and idler photons (with frequency ${\omega }_{i}$), respectively. $\Omega \equiv \omega_s-\omega_{p}={{\omega }_{p}}-{{\omega }_{i}}$ is the frequency detuning  of signal photons or idler photons. $f(\Omega )$ is the spectral amplitude of the biphoton state.

The signal and idler photons  are distributed to two parties, named Alice and Bob, over optical fibers. At Alice side, there is an object with a specific reflectivity pattern. The signal photons are firstly dispersed to different directions along a line according to their frequencies by a spatial dispersion component, such as a grating. Then they illuminate the object along a line, which is named as the illuminating line. Signal photons with different frequenciy will arrive at different positions on the object. The signal photons with a specific frequency detuning $\Omega$ would arrive at a specific position $x_{\Omega}$ in the line. The reflectivity pattern along the illuminating line would modulate the spectrum of the reflected signal photons. Hence, the positive-frequency field operator of the signal photons at Alice side can be expressed as 
{\setlength\arraycolsep{2pt}
\begin{eqnarray}
\hat{E}_s^{+}(t_s,L_s)&\sim&\int\mathrm{d}\Omega r(x_{\Omega})a_s(\omega_p+\Omega)\\ \nonumber
&&\quad\times e^{j(\omega_p+\Omega)t_s-j\left[\beta_{s0}+\beta_{s1}(\Omega-\Omega_0)\right]L_s},
\end{eqnarray}
where $a_s(\omega_p+\Omega)$ is the annihilation operator of the signal photons at the frequency $\omega_p+\Omega$. $t_s$ is the detection time of the signal photons, $L_s$ is the length of the optical fiber between the photon-pair source and the SPD at Alice side. Assuming that $L_s$ is small, the group velocity dispersion (GVD) in the fiber has been neglected. The phase coefficient of the signal photons in the fiber has been expanded in the vicinity of $\omega_p+\Omega_0$ , which is the central frequency of signal photons,  as $\beta_s(\omega_p+\Omega)=\beta_{s0}+\beta_{s1}(\Omega-\Omega_0)$.  

At Bob side, the idler photons are temporally dispersed before detected by a SPD. The most convenient way to realize the dispersion is utilizing the GVD of the transmission fiber, by which the idler photons are sent to Bob side from the source. If the length of the optical fiber between the photon-pair source and the SPD at Bob side is 
$L_i$, then the positive-field operator of the idler photons at time $t_i$ can be expressed as 
\begin{eqnarray}
\hat{E}_i^{+}(t_i,L_i)&\sim&\int{\mathrm{d}\Omega a_i(\omega_p-\Omega)e^{j(\omega_p-\Omega)t_i-j\beta_{i}L_i}},
\end{eqnarray}
where $a_i(\omega_p-\Omega)$  is the annihilation operator for idler photons at the frequency $\omega_p-\Omega$. The phase coefficient of the idler photons in the fiber can be expanded as
\begin{eqnarray}
\beta_i(\omega_p-\Omega)&=&\beta_{i0}+\beta_{i1}(\Omega_0-\Omega)+\frac{1}{2}\beta_{i2}(\Omega_0-\Omega)^2,
\end{eqnarray}
where $\omega_p-\Omega_0$ is the central frequency of idler photons. The term with $\beta_{i2}$ is corresponding to GVD. Higher order dispersions in the fiber have been neglected. 

The signal and idler photons are detected by SPDs at both sides with arrival time recorded. Then Alice sends the single photon events to Bob. Time discriminated coincidence measurement can be carried out at Bob side. It can be analyzed by the second-order Glauber correlation function $G^{(2)}$. Under the assumption that the temporal dispersion introduced at Bob side is large enough [see Section 1 of Supplement 1], the $G^{(2)}$ function can be expressed as  
\begin{eqnarray}
G^{(2)}(t_s,L_s;t_i,L_i)&=&\left\vert\langle 0\vert\hat{E}_i^{+}(t_i,L_i)\hat{E}_s^{+}(t_s,L_s)\vert\Psi\rangle\right\vert^2\\ \nonumber
&\sim&\left\vert f(\Omega)r(x_{\Omega})\right\vert^2_{\Omega=\tau/{\beta_{i2}L_i}},
\end{eqnarray}
where $\tau=(t_s-\beta_{s1}L_s)-(t_i-\beta_{i1}L_i-\beta_{i2}\Omega_0L_i)$.

According to equation (5), the coincidence measurement result has the profile of the spectrum of the biphoton state, which is modulated by the reflectivity pattern along the illuminating line in a nonlocal way [17,18].
 Hence, the reflectivity pattern $r(x_{\Omega})$ can be extracted from the coincidence measurement result. In this scheme, the signal photons are detected without discriminating their frequency and arrival time, hence, the reflectivity pattern along the illuminating line can not be recovered by Alice side only. On the other hand, the idler photons are discriminated according to their frequencies by the arrival time, supporting the nonlocal imaging process in the way of QGI.

The principle can be explained by the transformation of quantum correlation in this scheme. At Alice side, signal photons with different frequencies are mapped to different illuminating positions on the object due to spatial dispersion component. While, at Bob side, the idler photons with different frequencies have different arrival time due to the GVD of the fibers. Hence, the frequency correlation in the photon pairs is transformed to the correlation between the illuminating positions of signal photons on the object and the arrival time of idler photons.  Compared with QGI schemes based on momentum-momentum or position-position correlation[1] , our scheme realizes QGI in time domain thanks to the specific correlation. Since the telecom-band frequency correlated photon pairs can be easily distributed over optical fiber, this scheme provides a practical way to realize long distance QGI. 


\begin{figure}[htbp]
\centering
\fbox{\includegraphics{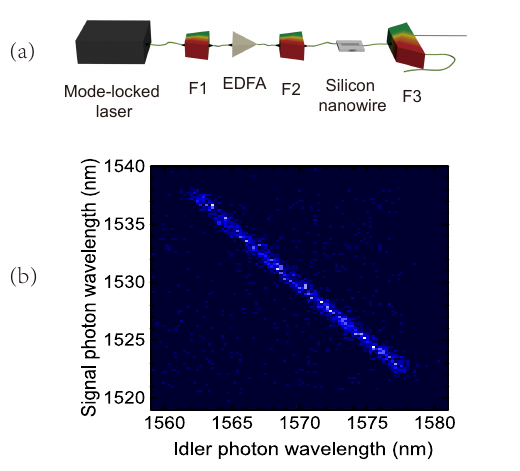}}
\caption{(a), The frequency correlated  photon-pair source based on the SFWM process in a silicon nanowire waveguide. F1, F2 and F3 are optical filters for pump and signal/idler photons. EDFA: Erbium doped fiber amplifier. (b), The joint spectral density of the biphoton state.}
\label{fig1}
\end{figure}
In the experiment, the frequency correlated photon pairs are generated by the SFWM process in a silicon nanowire waveguide [15]. The sketch of the source is shown in Fig. \ref{fig1}(a). The pulsed pump light has a center wavelength of $1550.92$ nm, a pulse width of about $3.7$ ps and a repetitive rate of $40$ MHz. It is generated by a mode locked laser, then a filter (F1) is used to confine its spectrum and extend its pulse width. An erbium doped fiber amplifier (EDFA) is used to increase the pump power and the filter (F2) after the EDFA is used to suppress the amplified spontaneous emission at the signal and idler wavelengths.  
The filters F1 and F2 are composed of dense wavelength division multiplexing devices (DWDMs). The pump light is injected into a silicon nanowire waveguide ($11$ mm in length, with a cross section of  $500$ nm $\times$ 220 nm) through a lensed fiber. Thanks to the high nonlinearity and the designed low dispersion property of the silicon nanowire waveguide, broad-band frequency correlated photon-pairs can be generated. The signal and idler photons are separated by a filter system (F3) composed of cascaded coarse wavelength division multiplexing devices (CWDMs). The center wavelengths of signal and idler photons are $1530$ nm and $1570$ nm, respectively.

The joint spectrum density (JSD) of the biphoton state is measured by the method shown in ref. [19]. The signal and idler photons are dispersed temporally by a piece of 50 km-long SMF. The measured JSD is shown in Fig. 1(b). It is shown that the wavelengths of  signal and idler photons are anti-correlated. The wavelength spans of the signal and idler photons are $16$ nm, which are determined by the bandwidth of the filters F3.

\begin{figure}[htbp]
\centering
\fbox{\includegraphics{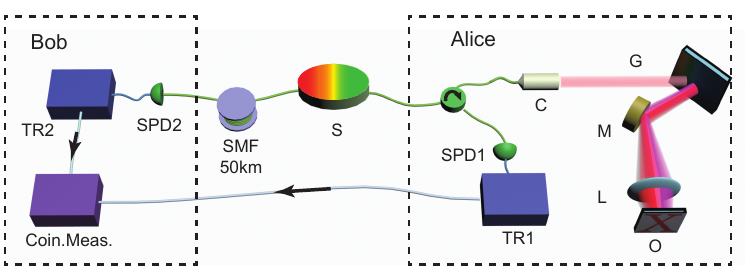}}
\caption{Experiment setup. The frequency correlated photon-pair source (S) has been shown in Fig. \ref{fig1}  The signal photons are distributed to Alice through a short piece of SMF and to Bob through 50 km-long SMF.  At Alice side, the signal photons are collimated to a spatial optical beam by a collimator (C) after an optical fiber circulator (CIR), then spatially dispersed by a grating (G), and illuminate the object (O) after a focal lens (L). At Bob side, the GVD of the transmission fiber is utilized to realize temporal dispersion. At both sides, photons are detected by SPDs (SPD1 \& SPD2), and the time of single photon events are recorded(time recorder, TR1 \& TR2), and coincidence measurement (Coin.Meas.) is carried out. M: mirror.}
\label{fig2}
\end{figure}

The sketch of the setup for long distance temporal QGI is shown in Fig. \ref{fig2}. The source shown in Fig. \ref{fig1} provides frequency correlated photon pairs. The signal photons are sent to Alice by a short  piece of SMF. At Alice side, the signal photons are delivered to an spectrally encoded confocal microscopy (SECM)[20,21] system through a circulator (CIR). In the SECM configuration, the photons are collimated to a spatial optical beam with a $1/e^2$ diameter of $2.1$ mm. A reflective diffraction grating ($600$ Line/mm, blaze wavelength equal to 1600nm) is used to disperse the signal photons spatially. After a lens with a focal length $l=25.4$ mm, the signal photons are focused on the surface of the object along a line, i.e., the illuminating line, with a length of about 250 \textmu m . The signal photons with different frequencies would be focused on different positions along the illuminating line. 

The object is a standard photolithographic mask, which is a silica plate with a patterned chrome layer. It provides a reflectivity pattern. The signal photons are partially reflected according the reflectivity pattern along the illuminating line, resulting in modulated spectrum. The reflected signal photons are collected by the collimator and sent to a SPD (ID220, ID Quantique) through the CIR. The single-photon events are recorded with high time-resolution using a time recoder (TR1) in a photon correlator (DPC 230, Becker \& Hickl GmbH).

While, the idler photons are sent to Bob through 50 km-long SMF. Then the idler photons are detected by the other SPD and the single photon events are also recorded with high time-resolution (TR2). On one hand, the frequencies of the idler photons are preserved after the propagation through such a long fiber, hence, the long distance distribution of the frequency correlated biphoton states is realized. On the other hand, the transmission fiber provides a GVD of $d=900$  ps/nm (estimated by the group velocity dispersion parameter of SMF). Since the bandwidth of idler photon is 16 nm, the idler photons are temporally broadened to $15.4$ ns.

When sufficient single photon events are collected at Alice and Bob sides, Alice should send the records of the detected signal photons to Bob by a classic channel, and Bob could realize coincidence measurement utilizing recorded signal and idler photon events. In the experiment, the single photon events at both sides are recorded by two channels of the photon correlator with a resolution of $164.61$ ps, and the coincidence is calculated by a computer according to the records. A typical measurement result is shown in Fig. \ref{fig3}(a), while the inset shows the reflectivity pattern of the object. The dimension of the pattern is 150 \textmu m x 140 \textmu m. The red areas are the regions with high reflectivity. The dashed line is the illuminating line corresponding to the histogram. It can be seen that the histogram has the shape of the pattern along the illuminating line clearly.

\begin{figure}[htbp]
\centering
\fbox{\includegraphics{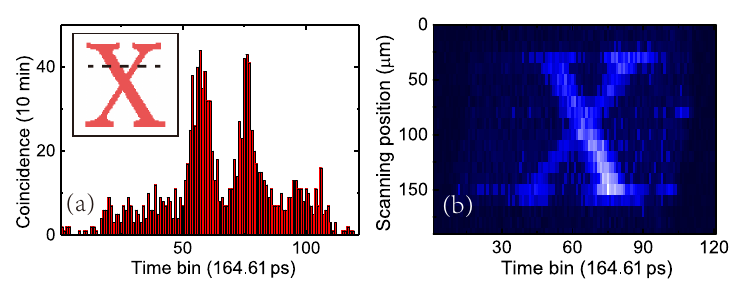}}
\caption{QGI of the object. (a), A typical coincidence histogram. The inset is the reflectivity pattern of the object and the dashed line is the illuminating line corresponding to the coincidence histogram. (b), The image obtained by step-moving the object with a step of 10 \textmu m. }
\label{fig3}
\end{figure}

By step-moving the object along the direction orthogonal to the illuminating line, two-dimensional imaging of the object can be obtained, which is shown in Fig.\ref{fig3}(b). The moving step is $10$~\textmu m. The measurement time for each illuminating line is 10 minutes. The coincidence counts in every time bin are indicated by different colors.  It can be seen that a clear image of the object has been obtained. Considering the long distance between Alice and Bob, this temporal QGI scheme realizes a function, of long distance image transmission like a ``quantum faxing machine''. It is worth noting that in the experiment the image generated by the temporal QGI has a little distortion. It is due to the fiber length variation when the temperature changes during the measurement. Since the fiber is as long as $50$ km, a temperature variation of  $1^\circ\mathrm{C}$ would lead to an arrival time variation of several nanoseconds for the idler photons traveling through the fiber.

To analyze the resolution of our QGI scheme, another object with narrower line width is used, as shown in Fig.\ref{fig4}(a). The image generated by the temporal QGI experiment is shown in Fig.\ref{fig4}(b). The line width of the object is 13 \textmu m at the dashed line, while the width of the image at the corresponding position is about 1.65 ns. According to the experiment parameters [see Section 2 of Supplement 1], it is related to a line width of 25.6 \textmu m, much wider than the actual width. The broadening of the line is due to the limited resolution of the experiment, which is mainly determined by the wavelength resolving power of the grating at Alice side and the timing jitters of the single photon detecting system (including the SPDs and the photon correlator).

\begin{figure}[htbp]
\centering
\fbox{\includegraphics{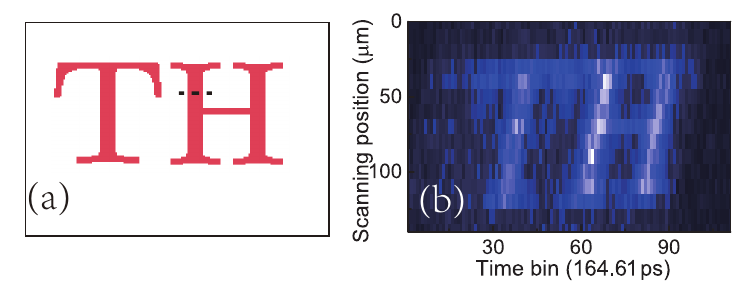}}
\caption{QGI using a object with narrower line width. (a), The reflectivity pattern of the target object. The line width is 13 um at the dashed line, while the width of the image at the corresponding postion is 25.6 \textmu m.  (b), The image generated by the temporal QGI experiment. }
\label{fig4}
\end{figure}

Along the illuminating line,  a reflectivity point at the object has a point-spread spectral function, which has a full width at half maximum (FWHM) determined by the resolving power of the grating [22],
\begin{eqnarray}
\delta\lambda&=&\frac{\lambda_{s0}}{N},
\end{eqnarray}
where $\lambda_{s0}=\frac{2\pi c}{(\omega_p+\Omega_0)}$ is the centeral wavelength of signal photons in vaccum, $N$ is the total number of grooves in the beam area on the grating. The point spread spectral function will result in  a point-spread coincidence peak, with a FWHM of 
\begin{eqnarray}
\delta\tau_1=\delta\lambda\times d,
\end{eqnarray}
where $d$ is the temporal dispersion at Bob side.
On the other hand, the impact of the single photon detecting system is denoted by $\delta\tau_2$, including the effect of timing jitters of  two SPDs and the time resolution of photon correlator. 
Hence, the resolution of the temporal QGI can be estimated by 
{\setlength\arraycolsep{2pt}
\begin{eqnarray}
\delta x&=&\frac{\sqrt{\delta\tau_1^2+\delta\tau_2^2}}{d}\cdot l\left.\frac{\mathrm{d}\theta}{\mathrm{d}\lambda}\right\vert_{\lambda=\lambda_{s0}}=\frac{l}{p\cos\theta_0}\sqrt{\frac{\lambda_{s0}^2}{N^2}+\frac{\delta\tau_2^2}{d^2}},
\end{eqnarray}
where, $\theta_0$ is the diffraction angle of the signal photons with wavelength $\lambda_{s0}$, $p$ is the period of the grating. The relation $p(\sin\theta-\sin\theta_i)=\lambda$ between the diffraction angle $\theta$ of the signal photons, the wavelength $\lambda$ and  the incident angle $\theta_i$, has been used in the equation.

According to the parameters of the experiment setup ($l=0.0254$ m, $p=1.67$ \textmu m, $\theta_0 ={{11.9}^{{}^\circ }}$, ${{\lambda }_{s0}}=1530$ nm, $N=1044$ grooves, $\delta {{\tau }_{2}}=389$ ps, $d=900$ ps/nm), the calculated spatial resolution is about 23.8 \textmu m, agrees well with the experiment results considering the fact that the reflectivity of the object has a rectangluar profile. 

Equation (7) shows that the spatial resolution could be improved through (1) increasing temporal dispersion at Bob side; (2) improving the time resolution of the single photon detection system; (3) increasing the beam diameter at Alice side; (4) utilizing an objective lens with smaller focal length at Alice side. Our recent work [23] shows that $\tau_2=80$ ps can be realized in the single photon detection system composed of super conducting nanowire single photon detectors (timing jitter $\sim60$ ps) and high performance photon correlator (timing jitter $\sim12$ ps). If the diameter of the signal photon beam is expended to $5$ mm and an objective lens with a focal length of $5$ mm is used, a resolution as high as $1.1$ \textmu m can be expected.

As a conclusion, we proposed and experimentally demonstrated a scheme of temporal  QGI. By step-move the object, the object can be imaged through 50 km SMF by this scheme. This scheme works like a “quantum fax machine” over optical fibers. The spatial resolution of this scheme is analyzed, showing that it could be less than the wavelength of the illuminating photons by proper system design. It will extend the application of QGI at large geographical scale.

\section*{Funding Information}
This work was supported by 973 Programs of China under Contract No. 2011CBA00303 and 2013CB328700, Tsinghua University Initiative Scientific Research Program, Basic Research Foundation of Tsinghua National Laboratory for Information Science and Technology (TNList).

\clearpage

\chapter*{Appendix}
\section*{S1. $G^{(2)}$ function}
It is well known that for biphoton state $\vert\Psi\rangle$, the second-order Glauber correlation function $G^{(2)}$ can be calculated using [1]
{\setlength\arraycolsep{2pt}
\begin{eqnarray}
G^{(2)}(t_s,L_s;t_i,L_i)&=&\left\vert\langle0\vert\hat{E}_s^{+}(t_s,L_s)\hat{E}_i^{+}(t_i,L_i)\vert\Psi\rangle\right\vert^2,
\end{eqnarray}
where $t_s$, $t_i$ is the detection time of signal and ider photons, $L_s$, $L_i$ is the optical path of signal and idler photons, respectively.
Including the dispersion of the transmission fiber, phases will be introduced to the positive-frequency operators, as shown in the Eq.(2) and Eq.(3), following which $\langle0\vert\hat{E}_s^{+}(t_s,L_s)\hat{E}_i^{+}(t_i,L_i)\vert\Psi\rangle$ can be calculated with the state in Eq.(1),
{\setlength\arraycolsep{2pt}
\begin{eqnarray}
&&\langle 0\vert \hat{E}_s^{+}(t_s,L_s)\hat{E}_i^{+}(t_i,L_i)\vert\Psi\rangle \nonumber\\
&\sim &\iiint\mathrm{d}\omega_s
\mathrm{d}\omega_i\mathrm{d}\Omega f(\Omega)r(x_{\Omega})e^{j(\omega_st_s-\beta_sL_s)+j(\omega_it_i-\beta_iL_i)}\nonumber\\
&&\times\langle0\vert a_s(\omega_s)\vert \omega_p+\Omega\rangle_s \langle0\vert a_i(\omega_i)\vert \omega_p-\Omega\rangle_i\\
&=&\iiint\mathrm{d}\omega_s
\mathrm{d}\omega_i\mathrm{d}\Omega f(\Omega)r(x_{\Omega})e^{j(\omega_st_s-\beta_sL_s)+j(\omega_it_i-\beta_iL_i)}\nonumber\\
&&\times\delta(\omega_p+\Omega-\omega_s) \delta(\omega_p-\Omega-\omega_i)\\
&=&e^{j\varphi}\int\mathrm{d}\Omega f(\Omega)r(x_{\Omega})e^{j\Omega\tau-j\beta_{i2}\Omega^2/2}\\
&=&{2\pi}e^{j\varphi}\mathcal{F}_\tau(f(\Omega)r(x_{\Omega}))*\mathcal{F}_\tau(e^{-j\beta_{i2}\Omega^2/2}),
\end{eqnarray}
where we used the phase coefficient expandation in the Eq.(2) and Eq.(4) and the commutation relation $[a_m(\omega_1), a_n^{\dagger}(\omega_2)]=\delta_{m,n}\delta(\omega_1-\omega_2)$, $m,n\in \{s,i\}$, $\delta_{m,n}$ is the Kronecker delta function, and $\delta(\cdot)$ is the Dirac delta function. $e^{j\varphi}$ includes all the $\Omega$-independent phases, and  $\tau=(t_s-\beta_{s1}L_s)-(t_i-\beta_{i1}L_i-\beta_{i2}\Omega_0L_i)$. $\mathcal{F}_{\tau}(\cdot)$ denotes the inverse Fourier transformation, as
\begin{eqnarray}
\mathcal{F}_\tau(f(\Omega)r(x_{\Omega}))&=&\frac{1}{2\pi}\int{\mathrm{d}\Omega f(\Omega)r(x_{\Omega})e^{j\Omega\tau}},\\
\mathcal{F}_\tau(e^{-j\beta_{i2}\Omega^2/2})&=&\frac{1}{2\pi}\int{\mathrm{d}\Omega e^{-j\beta_{i2}\Omega^2/2}e^{j\Omega\tau}}\nonumber\\
&=&\frac{1}{\sqrt{2\pi j\beta_2L_i}}e^{-j\tau^2/2\beta_{i2}L_i}.
\end{eqnarray}

Combining equation (13), (14) and (15), we obtain
\begin{eqnarray}
&&\langle 0\vert \hat{E}_s^{+}(t_s,L_s)\hat{E}_i^{+}(t_i,L_i)\vert\Psi\rangle\nonumber\\
&\sim&\int{\mathrm{d}\tau_1\left\{\int{\mathrm{d}\Omega f(\Omega)r(x_{\Omega})e^{j\Omega\tau_1}}\right\}e^{-j(\tau-\tau_1)^2/2\beta_{2i}L_i}}\\
&\sim&\int{\mathrm{d}\tau_1\left\{\int{\mathrm{d}\Omega f(\Omega)r(x_{\Omega})e^{j\Omega\tau_1}}\right\}e^{j\frac{\tau\tau_1}{\beta_{i2}L_i}-j\frac{\tau^2_1}{2\beta_{i2}L_i}}}\\
&\sim&f(\Omega)r(x_{\Omega})\vert_{\Omega=\frac{\tau}{\beta_{i2}L_i}}.
\end{eqnarray}
From Eq.(17) to Eq.(S18), we have assumed that a very large dispersion has been introduced at Bob side, so that $\beta_{i2}L_i$ is much larger than the temporal width of idler-photon wavepackets, and neglected the term  $\frac{\tau_1^2}{2\beta_{i2}L_i}$, because $\frac{\tau_1^2}{2\beta_{i2}L_i}\ll 1$ [2] .

Lingking Eq.(9) and Eq.(18) for clearity, we can get the expression of $G^{(2)}$ function when large temporal dispersion is introduced at Bob side,
\begin{eqnarray}
G^{(2)}(t_s,L_s;t_i,L_i)&\sim&\left\vert f(\Omega)r(x_{\Omega})\right\vert^2_{\Omega=\tau/{\beta_{i2}L_i}},
\end{eqnarray}
where $\tau=(t_s-\beta_{s1}L_s)-(t_i-\beta_{i1}L_i-\beta_{i2}\Omega_0L_i)$.

According to Eq.(19), on one hand, it is indicated that the coincidence measurement results have the shape of the spectrum of the biphoton state, which is nonlocally modulated by the reflectivity pattern of the object, so by extracting the reflectivity spectra $r(x_{\Omega})$, the image of the  object can be reconstructed.
On the other hand, it can be seen that, after temporal dispersion, the coincidence peak is spread to have a width of  $\tau_w=\beta_{i2}L_i\Omega_w$, where $\Omega_w$ is the spectrum width of the biphoton state.  The spreading of the coincidence peak is due to the GVD introduced at Bob side. For idler photons, using the expandation of the phase coefficient in Eq.(4), the group velocity of idler photons can be expressed as [2]
\begin{eqnarray}
\frac{1}{v_{ig}}=\frac{\partial \beta_i}{\partial \omega_i}=-\frac{\partial \beta_i}{\partial \Omega}=\beta_{i1}+\beta_{i2}(\Omega_0-\Omega).
\end{eqnarray}
So the time needed for idler photons from the source to the SPD at Bob side will be 
\begin{eqnarray}
t_{\Omega}=\frac{L_i}{v_{ig}}=\beta_{i1}L_i+\beta_{i2}L_i(\Omega_0-\Omega).
\end{eqnarray}
For idler photons with frequency difference $\Delta\Omega$, the time delay difference will be 
\begin{eqnarray}
\Delta t=-\beta_{i2}L_i\Delta\Omega,
\end{eqnarray}
which is coincident with the spread of the coincidence peak.
In fact, after the spatial dispersion at Alice side, and the temporal dispersion at Bob side, the frequency correlation in the photon pairs is transformed to the correlation between the illuminating position $x_\Omega$ of signal photons on the object and the travel time $t_\Omega$ of idler photons before detected by the SPD. Compared with previous QGI schemes based on momentum-momentum or position-position correlation, our time domain QGI is realized utilizing this kind of correlation.

\section*{S2. Scale transformation relation between the object and the time-domain image}
This section discusses the relation between the scale of the object and the image.
It is clear that in the direction orthogonal to the illumiating line, the scale of the image should be equal to the scale of the object. In the following, we focus on the scale transformation relation in the direction of the illuminating line.

At Alice side, the relation between the illuminating position $x_{\Omega}$ on the object and the signal photon frequency $\omega_p+\Omega$  can be expressed as 
\begin{eqnarray}
x_\Omega=\left. x_{\Omega_0}-2\pi cl\frac{\Omega-\Omega_0}{\left(\omega_p+\Omega_0\right)^2}\frac{\mathrm{d}\theta}{\mathrm{d}\lambda}\right\vert_{\lambda=2\pi c/(\omega_p+\Omega_0)},
\end{eqnarray}
where $x_{\Omega_0}$ is the illuminating position of signal photons with frequency $\omega_p+\Omega_0$(the central frequency of the signal photons), $l$ is the focal length of the lens, $c$ is the velocity of light in vacuum, $\theta$ is the diffraction angle of light beams after grating, and $\lambda$ is the wavelength of signal photons, which is corresponding to the frequency by the relation $\lambda=2\pi c/(\omega_p+\Omega)$.
As we use the first order of interference of the grating, the angle of the incident light $\theta_i$, the angle  $\theta$ of the diffration light with wavelength $\lambda$, must satisfy the equation [3], $p\left(\sin\theta-\sin\theta_i\right)=\lambda$, where  $p$ is the period of the grating. Hence,
\begin{eqnarray}
\left\vert\frac{\mathrm{d}\theta}{\mathrm{d}\lambda}\right\vert_{\lambda=\frac{2\pi c}{\omega_p+\Omega_0}}&=&\frac{1}{p\cos\theta_0},
\end{eqnarray}
where, $\theta_0$ is the diffraction angle of signal photons with frequency $\omega_p+\Omega_0$.
Two points with high reflectivity on the object with distance $\Delta x$ will reflect signal photons with frequency differency equal to  
\begin{eqnarray}
\Delta \Omega=-\frac{\left(\omega_p+\Omega_0\right)^2}{2\pi cl}p\cos\theta_0\Delta x.
\end{eqnarray}
where, $\theta_0$ is the diffraction angle of signal photons with frequency $\omega_p+\Omega_0$.

According Eq.(22), the time differency between the two coincidence peaks resulted from the two high-reflectivity points will be
\begin{eqnarray}
\Delta \tau=-\beta_{i2}L_i\Delta\Omega=\frac{\left(\omega_p+\Omega_0\right)^2}{2\pi cl}\beta_{i2}L_ip\cos\theta_0\Delta x.
\end{eqnarray}

Considering the relation between the dispersion parameter of $d$ and $\beta_{i2}$, i.e. $d=-\frac{2\pi c}{\lambda^2_{i0}}\beta_{i2}L_i$, $\lambda_{i0}=2\pi c/(\omega_p-\Omega_0)$ is the central wavelength of idler photons, equation (26) can be expressed using the dispersion parameter $d$ as 
\begin{eqnarray}
\Delta \tau&=&-\frac{\lambda_{i0}^2}{\lambda_{s0}^2l}pd\cos\theta_0\Delta x,
\end{eqnarray}
where $\lambda_{s0}$ is the central wavelength of signal photons, and  $\lambda_{s0}=2\pi c/(\omega_p+\Omega_0)$.

\end{document}